# The lunar surface as a recorder of astrophysical processes[1]

Ian A. Crawford[a,b], Katherine H. Joy[c], Jan H. Pasckert[d], Harald Hiesinger[d]

[a]Department of Earth and Planetary Sciences, Birkbeck College, University of London, Malet Street, London WC1E 7HX, UK. Orchid ID: 0000-0001-5661-7403

[b]Centre for Planetary Sciences at UCL/Birkbeck, Gower Street, London WC1E 6BT

[c]Department of Earth and Environmental Sciences, The University of Manchester, Oxford Road, M13 9PL, Manchester, UK. Orchid ID: 0000-0003-4992-8750

[d]Institut für Planetologie, Westfälische Wilhelms-Universität, Wilhelm-Klemm-Str. 10, 48149 Münster, Germany. Orchid IDs: 0000-0002-1338-5638, 0000-0001-7688-1965

# Summary

The lunar surface has been exposed to the space environment for billions of years and during this time has accumulated records of a wide range of astrophysical phenomena. These include solar wind particles and the cosmogenic products of solar particle events which preserve a record of the past evolution of the Sun, and cosmogenic nuclides produced by high-energy galactic cosmic rays which potentially record the galactic environment of the Solar System through time. The lunar surface may also have accreted material from the local interstellar medium, including supernova ejecta and material from interstellar clouds encountered by the Solar System in the past. Owing to the Moon's relatively low level of geological activity, absence of an atmosphere, and, for much of its history, lack of a magnetic field, the lunar surface is ideally suited to collect these astronomical records. Moreover, the Moon exhibits geological processes able to bury and thus both preserve and 'time-stamp' these records, although gaining access to them is likely to require a significant scientific infrastructure on the lunar surface.

# 1. Introduction

There are multiple scientific reasons for wishing to resume the robotic and human exploration of the lunar surface, ranging from lunar geology to astrobiology (for reviews see [1-4]). Other papers in this volume are mostly concerned with the potential of the lunar surface as a platform for astronomical observations of various kinds, whereas in this contribution we argue that the lunar surface itself will have recorded much of astrophysical interest. In this sense, the Moon itself can be viewed as a giant 'telescope' which has been observing and recording astrophysical processes ever since it first developed a solid surface some 4.5 billion years ago.

---

Several factors combine to make the lunar surface an ideal, and perhaps unique, recorder of a wide range of astrophysical processes throughout Solar System history. Primarily, this is because the lunar surface, unprotected by an atmosphere or, for much of its history, a magnetic field, has been directly exposed to the space environment for most of the last 4.5 Gyr. As a consequence, particles and radiation from space have impacted the lunar surface unimpeded, leaving evidence of their presence in the rocks and soils of the lunar regolith. Examples include solar wind particles and cosmogenic products of solar energetic particle events, and thus a record of the past evolution of the Sun, and cosmogenic nuclides produced by galactic cosmic rays, and therefore a record of the past galactic environment of the Solar System. The lunar surface may also have accreted material from the local interstellar medium, including supernova ejecta and material from interstellar clouds encountered by the Solar System in its journey around the Galaxy. Equally as important as the collection of these astrophysical records, however, are lunar geological processes which facilitate their long-term preservation. As a relatively low-mass planetary body, whose own internal geological processes largely ceased billions of years ago, the Moon has preserved an ancient surface with some sampled crustal rocks dating from 4.3-4.4 Gyr [4]. Crucially, and unlike the possibly equally ancient surfaces of some asteroids, for most of its history the Moon has been sufficiently geologically active to bury, preserve, and 'time-stamp' ancient astrophysical records in near-surface rocks and soils. Key processes include the covering of old surfaces by lava flows, pyroclastic deposits and impact crater ejecta blankets, and ancient records preserved by these processes may be recoverable by future space missions [1,5].

## 2. Lunar records of solar activity

Our knowledge of the past evolution of the Sun comes mainly from theoretical modelling (e.g. [6,7]) and observations of other solar-type stars having a range of ages (e.g. [8,9]). These studies indicate that, whereas the overall solar luminosity was probably only ~70% of its present value when the Sun formed, its faster rotation would have resulted in greatly enhanced magnetic activity and associated solar wind and UV and X-ray emission. Thus, we expect the luminosity of the Sun to have increased, and the strength of the solar wind and high-energy photon and particle emission to have decreased, throughout Solar System history. Both of these effects will have had implications for the habitability of the terrestrial planets and, in particular, for the environment within which life originated and evolved on Earth [10]. In addition, the decrease in solar magnetic activity has likely resulted in a corresponding increase in the galactic cosmic ray (GCR) flux in the inner Solar System owing to the shrinkage of the heliosphere [11], which may also have had consequences for biological evolution on Earth.

Although generally accepted, the low total luminosity of the Sun in its early history is difficult to understand given the evidence for liquid water on the surfaces of early Earth and Mars (i.e. the 'faint-young-Sun' paradox [12]). Most proposed explanations invoke enhanced greenhouse gas concentrations in the atmospheres of these planets, although difficulties remain with these models [13]. One suggested alternative explanation is that the Sun may have been more massive, and thus more luminous, in the past, but that it lost several percent of its initial mass in strong solar winds early in its history [14,15]. Observations of limited mass-loss from young solar-type stars have cast doubt on this proposal [16], but direct measurements of the strength of the solar wind through time could in principle settle the issue. In any case, it is clear that obtaining direct observational evidence of solar activity through time would not only provide important insights into the evolution of the Sun as a star, but also improve our understanding of the evolution of the atmospheres, surface environments and habitability of the terrestrial planets. Here we argue that the near-surface environment of the Moon has the potential to provide this valuable information.

The Sun is an emitter of both low and high energy particles which may potentially yield information about solar processes and evolution. Solar wind particles (electrons and protons, alpha particles, and trace heavy ions with energies up to ~10 MeV) are emitted constantly from the Sun's corona, varying in intensity with solar activity. Higher energy (~10 to $10^3$ MeV) particles, often referred to as solar cosmic rays (SCRs) or solar energetic particles (SEPs), are episodically emitted during solar flares and coronal mass ejection events and are able to produce a range of cosmogenic nuclides when they impact planetary surfaces (e.g. [17-19]).

Analyses of samples returned by the Apollo and Luna missions have revealed that the lunar regolith is an efficient collector of solar wind particles and cosmogenic nuclides produced by SCRs [5,17,20], and that it therefore potentially contains a record of past solar activity (e.g. [21-24]). In this context, determining the time dependence of both the flux and composition of the solar wind would be of interest. Whereas the overall solar wind flux is probably the most direct indicator of solar activity, in practice the surfaces of regolith particles can become saturated with solar wind [25], and the solar wind concentration retained in the regolith is influenced by each individual grain's exposure history [e.g. 25-29].These factors may make regolith particles insensitive to recording bulk temporal flux variations. However, there is evidence that changes in solar activity also affect the relative abundances of ions in the solar wind owing to differential ionisation in the solar wind source regions [30], potentially making the *composition* of the solar wind a proxy for solar activity. In addition, early solar activity, and especially the frequency of coronal mass ejection events, may have been responsible for the wholesale depletion of moderately volatile elements such as Na and K in the surficial regolith, which may also detectable in lunar samples [31].

The implantation depths of solar wind ions into regolith particles depend on the irradiation energy, the mass of the irradiating particles, and the composition (chemistry and mineral lattice structure) of the target material (e.g. [32,33]). To access these records, research has focused on determination of the light element (H, C, O, N) and noble gas isotope (He, Ne, Ar, Kr, Xe) budgets of small rock fragments or individual mineral grains (e.g. [25,34,35]), and analysis of depth-dependent concentrations of these elements within the grains (e.g. [36]) utilising noble gas acid-step leaching techniques (e.g. [37]) and secondary ion microprobe analyses (e.g. [38]). However, all these efforts are constrained by the nature of the existing lunar sample collection. Of necessity, samples collected by the Apollo and Luna missions (see [39] for a review) were obtained from the present-day surface of the Moon and most have had a very long, but generally indeterminate, exposure to the solar wind. Moreover, unless they have been deeply buried and recently exhumed, surface samples are unlikely to have sampled the most ancient (i.e. several Gyr-old) solar wind that is of greatest interest in investigating early solar evolution, although progress may be made by studying solar wind trapped in ancient regolith breccias dating from that time [40]. Similar considerations apply to inferring past solar activity from cosmogenic nuclides produced by SCRs (e.g. [17,41]).

A key requirement for further progress would be to obtain independent information on the absolute ages of both the start and end times of solar wind and SCR exposure for a range of samples exposed to the space environment at widely different times in the past. Fortunately, just such 'time-stamped' samples likely exist on the Moon in the form of ancient regoliths (hereinafter 'palaeoregoliths') that were exposed to the solar wind and

SCRs at discrete times in the past and then covered up, and thus preserved, by later geological processes. Obtaining such samples would greatly help in reconstructing a record of solar activity through time (although the individual grain exposure histories within sampled palaeoregolith deposits will still need to be considered).

Before leaving this discussion of lunar records of past solar activity, we draw attention to the possibility that the vertical temperature profile in the uppermost few metres of the lunar regolith may record the history of the solar irradiance over the last several centuries [42]. Such measurements have been applied to reconstructing terrestrial surface temperature changes over similar timescales [43], but implementation on the Moon would yield a measure of solar irradiance variations free of the complexities introduced by Earth's atmosphere and climate. Building on results from the Apollo heat flow experiments, Miyahara et al. [42] calculated that a temperature measurement precision of ~0.01°C over a depth range of ~10 m would be able to distinguish between different models of the total solar irradiance back to the Maunder Minimum in the mid- to late-seventeenth century. In addition to providing information on very recent solar activity, such measurements may be helpful in understanding the historical evolution of Earth's climate system. Obtaining them will require the drilling of multiple boreholes to ~10 m depth (i.e. five times the depths of the Apollo heat-flow measurements of ~2 m).

## 3. Lunar records of galactic processes

As the Solar System has been orbiting the Galaxy for the last 4.6 Gyr it will have experienced a wide range of different galactic environments. The recent review of galactic rotation constants provided by Valleé [44] implies that the Sun traverses the entire spiral pattern of the Galaxy every ~720 to 1760 Myr (where the uncertainty arises from continuing uncertainties in the angular velocity of the spiral arms). As the Galaxy appears to have four major spiral arms [45], this implies spiral arm passages every ~180 to 440 Myr, during which periods the Solar System may have experienced a range of interesting astrophysical phenomena including nearby supernova (SN) explosions and transits through dense interstellar clouds. Reconstructing this history would provide astronomically valuable information on the structure and evolution of the Galaxy, as well as astrobiologically important information relevant for understanding the past habitability of our own planet [46-49]. Previous attempts to find correlations between spiral arm crossings and Earth's climate and extinction records have been controversial and ambiguous (e.g. [47]), in part be due to a lack of reliable geological records of the Solar System's astrophysical environment.

As reviewed in earlier publications [50,51], the lunar surface is likely to be a much better repository of this information for the same reasons that it will have preserved a record of ancient solar activity – i.e., it has been constantly exposed to the space environment throughout Solar System history, while also manifesting geological processes able to preserve records of this exposure. There are at least three forms that such records might take:

- Variations in the Galactic Cosmic Ray (GCR) flux, as recorded in the abundances of cosmogenic nuclides and/or radiation damage preserved in lunar surface rocks and soils
- Direct accretion of interstellar matter and/or supernova ejecta onto the lunar surface
- Variations in the lunar cratering rate driven by gravitational perturbations of the orbits of comets and/or asteroids by changes in the galactic gravitational environment.

## (a) Recording variations in GCR Flux

Several galactic processes affect the GCR flux in the inner Solar System, operating on a range of timescales [52-54]. On the longest timescales (>1 Gyr), the average GCR flux may reflect the galactic star formation rate, which could provide useful constraints on models of galactic evolution (although it would be necessary to account for an expected secular increase in GCR flux reaching the inner Solar System due to decreasing solar activity [11]). On timescales of the order of a few 100 Myr, the GCR flux is likely to be moderated by an enhanced supernova (SN) rate, and/or collapses of the heliosphere owing to encounters with dense interstellar clouds, associated with the Sun passing through galactic spiral arms [55-57]. On still shorter timescales (tens of Myr), additional variations in the GCR flux may be expected owing to the oscillation of the Sun about the plane of the Galaxy (with a period of ~64 Myr and amplitude ~70 parsecs [57]), and possible short-term variations in the size of the heliosphere owing to fluctuations in the local interstellar medium density [11]. Stochastic events, such as nearby (say, ≤50 parsecs) SN explosions, may be superimposed on these secular and quasi-periodic galactic influences. For example, Melott et al. [58] have considered the GCR flux at the Earth due to a SN at distance of 50 parsecs and find that the GCR flux could be elevated by between one and three orders of magnitude above its current value for several thousand years (see figure 1 in ref. [58]).

There are at least two ways in which evidence for GCR variations might be recovered from exposed Solar System samples. Firstly, the high energy GCR particles leave tracks of radiation damage in exposed materials, the density of which is proportional to the GCR flux and exposure duration ([5,22,59]). Secondly, when GCRs interact with atomic nuclei in geological materials a variety of cosmogenic nuclides (e.g. $^3$He, $^{10}$Be, $^{21}$Ne, $^{36}$Cl, $^{38}$Ar) are produced as a result of spallation and neutron capture reactions (e.g. [60]). Typically, these interactions occur within the uppermost metre or so of the exposed surface (e.g. [5,61]). Results of searches for GCR variations based on meteorite samples have proved to be controversial and inconclusive (e.g. [62,63]), in part because they only record an integrated GCR flux since becoming exposed to the space environment. As Wieler et al. [62] note "because of the limited sensitivity of the time-integrated GCR signals provided by meteorites, it is wise to consider … also the differential GCR flux signals provided by terrestrial sediment samples." Because terrestrial samples can be dated independently of the cosmic ray flux, this is a potentially powerful approach, but is limited by the relatively recent ages of terrestrial sedimentary samples, the complexity of Earth's geological and erosional history, and by the fact that the primary GCR flux is attenuated by the Earth's atmosphere and magnetic field. It is here that the lunar geological record may be able to help.

Several cosmogenic nuclides have been measured in lunar samples (e.g. [22,29,33,64]), and in principle the GCR flux could be inferred by measuring the density of cosmic ray tracks and/or the concentrations of cosmogenic nuclides in exposed lunar materials. In practice there are a number of complications, especially the generally unknown exposure and shielding histories (i.e. burial depths) of existing lunar samples. These uncertainties would be mitigated if the start and end times of the exposure of a given lunar sample, together with its burial history, could be determined independently. This will likely be key to reconstructing GCR records from which the changing galactic environment of the Solar System might be inferred. As for the solar wind history, the recovery of GCR records from buried palaeoregolith layers would be one possibility, although unlike the solar wind case the development of a surficial regolith may not be required to preserve GCR records because they will also occur at ~ metre depths within crystalline rocks.

We note in passing that a nearby SN explosion would also produce an enhancement in the neutrino flux, which in principle might be detected by damage tracks produced in mineral lattices by nuclei recoiling from a neutrino interaction. It has been proposed to search for such signals in terrestrial rock samples [65], but lunar samples would provide a longer temporal baseline and avoid the neutrino background produced in Earth's atmosphere.[2]

**(b) Recording the direct accretion of interstellar matter**

In its journey around the Galaxy, the Solar System will have been exposed to a range of different interstellar medium densities. As reviewed elsewhere ([66] and references cited therein), at present the Solar System appears to be located close to the boundary of a low density ($n_H \sim 0.1$-$0.2$ cm$^{-3}$, where $n_H$ is the density of hydrogen nuclei) interstellar cloud (the 'Local Interstellar Cloud', LIC), which is itself located in the even lower density ($n_H \sim 0.005$ cm$^{-3}$) and ~100 parsec radius Local Bubble within the Local ('Orion') Arm of the Galaxy. As noted by Cohen et al. [11], even small changes in the extent of the heliosphere caused by the Solar System moving in and out of low-density clouds like the LIC may have produced variations in the inner Solar System GCR flux on Myr timescales. On longer timescales, and especially during spiral arm passages, much denser interstellar environments are likely to be encountered, possibly resulting in the direct accretion of interstellar gas and dust onto the atmospheres and surfaces of the terrestrial planets [67-70].

In their study of the interaction of the Solar System with interstellar clouds, Yeghikyan & Fahr [68] found that for interstellar densities of $n_H \geq 1000$ cm$^{-3}$ the size of the heliosphere would shrink to less than one astronomical unit, leaving the Earth (and the Moon) directly exposed to interstellar material. The frequency with which the Solar System encounters such dense interstellar clouds is uncertain, with estimates varying between ~300 Myr and 3 Gyr [68,70-72]. For both astronomical and astrobiological reasons it would be desirable to reduce this uncertainty, and to determine whether or not encounters with dense interstellar clouds have influenced life on Earth. Pavlov et al. [67] calculated that if it took 200,000 years to cross a cloud with $n_H \sim 1000$ cm$^{-3}$ then ~1 kg m$^{-2}$ of interstellar dust would be deposited on exposed planetary surfaces, where it might be identified by distinctive chemical and isotopic signatures. Given the likely ages and relatively short durations of interstellar cloud encounters, the ancient and relatively undisturbed surface of the Moon appears far more likely to retain a record of such events than the dynamic surface of the Earth, especially if they have been preserved within independently dateable palaeoregolith deposits.

In addition to collecting interstellar dust, it is also possible that airless surfaces such as that of the Moon will provide a record of the gaseous component of interstellar clouds through which the Solar System has passed. One possibility would be interstellar pick-up ions, ionized and accelerated within the heliosphere and then implanted into the surfaces of lunar regolith grains [73]. In addition, now that there is abundant evidence for volatiles trapped in permanently shadowed regions (PSRs) at the lunar poles [74], where temperatures are typically of the order of 40 K [75], it may be worth considering whether directly accreted interstellar gas penetrating the inner heliosphere during interstellar cloud traverses could have become cold-trapped onto PSR surfaces; this might be a fruitful topic for future investigation.

---

[2] We thank Joe Silk for this observation.

As reviewed elsewhere [51], another component of interstellar material that might be identified on the lunar surface would be ejecta from nearby SN explosions. There has been a long-standing recognition that SN occurring within a few tens of parsecs, and perhaps as distant as 100 parsecs, may deposit debris enriched in radioactive elements within the Solar System (e.g. [76-79]), and evidence for two such events, in the age ranges of ~2 and ~8 Myr, has been reported from $^{60}$Fe deposition in ocean sediments [80-82]. Cook et al. [83] argued that the lunar surface has some advantages as a collector of SN ejecta as the much slower rate of surface re-working would allow it to accumulate in more concentrated layers than on Earth, and in 2016 this group [84] identified $^{60}$Fe enhancements in Apollo 12, 15 and 16 soil samples (collected from depths ≤few cm) consistent with the ~2 Myr-old SN event recognized in Earth ocean sediments. However, where the lunar record is likely to come into its own is in identifying debris from much older SN events than are likely to be preserved by Earth's dynamic surface environment. It is true that this will be complicated by the short half-lives ($T_{1/2}$ ≤ a few Myr) of radioisotopes likely to be present in SN ejecta (see table 1 of Fry et al. [79] for a summary), but two such isotopes, $^{146}$Sm ($T_{1/2}$ = 100 Myr) and $^{244}$Pu ($T_{1/2}$=80 My), are sufficiently long-lived to have recorded one or more spiral arm passages. In addition, careful analysis of the decay products of once-live isotopes in SN ejecta (e.g. $^{26}$Al, $^{53}$Mn, $^{60}$Fe, $^{41}$Ca) might also reveal the signatures of ancient SN events. We also note that Siraj and Loeb [85] have recently suggested that SN-accelerated dust grains might leave detectable tracks in mineral surfaces exposed at the lunar surface. Any such detections of SN ejecta would be expected to correlate with evidence for enhanced GCR (and neutrino) fluxes, so these different lines of evidence for ancient SN events would be mutually supportive.

Finally, we briefly mention an even more exotic possibility: recently it has been proposed that geological materials may record interactions with some candidates for dark matter particles [86] and, should the Solar System have encountered variations in the density of such particles in its orbit around the Galaxy, the long-lived lunar geological record would appear ideally suited to recording them.

### (c) Recording variations in impact cratering rate

The third category of lunar geological records with the potential to provide insights into galactic processes concerns the changing gravitational environment of the Solar System. It has long been speculated that the changing gravitational potential as the Solar System oscillates above and below the galactic plane, and passes through galactic spiral arms, may perturb the orbits of comets in the Oort Cloud and increase the impact cratering rate in the inner Solar System (e.g. [46,49,55,87-89]). Identifying possible periodicities and/or episodic spikes in the impact cratering rate, which might then be correlated with the Solar System's galactic environment, is problematical primarily because the terrestrial impact record is so sparse [90]. In contrast, the lunar surface holds an essentially complete impact record for most of Solar System history [91]. It follows that, by obtaining and dating samples of impact melt from a sufficiently large number (possibly hundreds) of lunar craters, it ought to be possible to determine unambiguously whether or not temporal variations in the impact flux have occurred and are correlated with galactic structure (although if most of the impactors were from the asteroid belt rather than from comets [92] any galactic signal might be muted). Note that we would expect any galactic periodicity in cratering rate to be correlated with the GCR flux, which is also recorded on the Moon. Obtaining such a complete impact record for the Moon would also have many other benefits for planetary science, including refining the inner Solar System impact cratering chronology and constraining models of the dynamical evolution of planetary orbits (e.g. [4] and references cited therein).

# 4. Preserving the record

The Moon will only record a history of past astrophysical processes if evidence for them has reached the surface and then been preserved by lunar geological processes. Before turning to the means of preservation, it is important to consider whether the lunar surface has always been as open to external influences as it is today. There are at least two aspects to consider: the ancient lunar magnetic field, and a possible early atmosphere.

Palaeomagnetic studies of Apollo samples suggest that between ~4.2 and ~3.6 Gyr ago the Moon had a core-generated magnetic field comparably strong to the Earth's present-day magnetic field (i.e. several tens of $\mu$T), which then declined by at least an order of magnitude prior to 3.2 Gyr ago [93]; it may have persisted at this reduced level (~5 $\mu$T) until about 2.5 Gyr ago, finally ceasing (<0.1 $\mu$T) sometime before ~1 Gr ago [94]. As a consequence, at least during the early part of this time period, the Moon's surface would have been partially shielded from the solar wind, so this would need to be taken into account in interpreting the most ancient solar wind records; possibly samples collected close to palaeomagnetic poles, where the magnetic field lines would tend to channel charged particles to the surface, would be preferred for such studies. On the other hand, high-energy GCRs, and uncharged particulate material (e.g. SN ejecta and interstellar dust particles) would not be expected to be significantly affected by an early lunar magnetic field. The presence of an ancient lunar atmosphere would potentially impede a wider range of exogenous material from reaching the surface, including some fraction of the primary GCRs. Given the Moon's low gravity, its only realistic opportunity to accumulate an atmosphere would be during periods of intense volcanic activity when the rate of magmatic degassing might transiently exceed the rate of atmospheric loss. Based on these arguments, Needham and Kring [95] estimated that a transient lunar atmosphere having a surface pressure of up to ~9 mbar (i.e. 1.5 times higher than the current atmospheric pressure on Mars) might have persisted for ~70 million years at the peak of mare volcanism. On the other hand, Wilson et al. [96] have argued that the intervals between individual mare eruptions (~20,000 to 60,000 years) would have been too long for such a transient atmosphere to accumulate. In any case, it appears that any ancient lunar atmosphere would only have affected the accumulation of astronomical records on the Moon's surface for geologically brief periods ~3.5 Gy ago.

Assuming that they reach the lunar surface unimpeded, the various astronomical records discussed in this paper will only have survived in detectable quantities if geological processes have acted to preserved them. This is especially true of material deposited directly onto the surficial regolith (e.g. solar wind particles, interstellar pick-up ions, and accreted interstellar material) because otherwise these records will be disturbed and diluted by the continuous comminution and overturning ("gardening") of the regolith by the unremitting impact of small meteoroids [20]. Owing to the deeper (order metre) penetration depths of GCRs, the record of cosmogenic nuclide formation, required to reconstruct variations in the GCR flux, is probably less sensitive to surficial regolith processing, but any record dating back hundreds of millions of years will still need to be protected from the disturbing effects of larger meteorite impacts. To illustrate this point, consider the solar wind and cosmogenic nuclei extracted from surface regolith samples collected by the Apollo missions. Solar wind and cosmogenic noble gases have been extracted from regolith samples collected at all six landing sites (see [33] for a recent review). However, these regoliths have mostly been developed on surfaces with ages exceeding 3 Gyr (e.g. [97]) so any evidence for temporal variations within them will have been smeared out by impact-induced gardening during these vast spans of time. Some exceptions are provided by samples collected on the ejecta blankets of young impact craters, such as South Ray crater (Apollo 16; estimated age 2 Myr), Cone crater (Apollo

14; 25 Myr), and North Ray crater (Apollo 16; 53 Myr), but these estimated ages [91] are mostly derived from cosmic ray exposure, so are not independent of the GCR fluxes that we seek to determine.

What we really need is to identify materials (e.g. palaeoregoliths) that were once exposed at the lunar surface for a known duration and which were subsequently covered by overlying material so that they have been isolated from the space environment ever since. Fortunately, there are at least three geological processes that will have acted to cover, and therefore preserve, pre-existing surfaces throughout lunar history:

- Eruption of low-viscosity basaltic lava flows
- Deposition of pyroclastic deposits around sites of explosive volcanism
- Emplacement of impact crater ejecta blankets

There are pros and cons associated with all three preservation mechanisms, which we now discuss.

**(a) Lava flows**

Basaltic lava flows cover ~17% of the lunar surface, mostly on the nearside, and their generally low viscosity and apparently laminar flow [98,99] suggests that palaeoregolith layers may be preserved beneath or between them. Once sampled, basalts can be dated to high accuracy using standard radiometric techniques (e.g. [97]), so the ages of palaeoregoliths trapped between lava flows could in principle be well constrained [5]. However, most mare basalts appear to have been erupted within the relatively narrow time interval between about 3.8 and 3.3 Gyr [100] (although older lava flows may be buried by younger ones now exposed at the surface), and there is no evidence for large-scale basaltic volcanism more recently than ~1 Gyr ago. It follows that palaeoregoliths dating from the last ~1 Gyr, which more or less encompasses the Solar System's most recent traverse of the galactic disk, are unlikely to be preserved between lunar lava flows unless younger, smaller scale, eruptions have occurred. In this context, it is worth noting that small patches of basaltic lavas, apparently erupted within the last 100 Myr, may have been identified on orbital imagery [101]. However, although these localities could potentially preserve valuable astronomical records from an important time interval, such young lavas are unexpected on the basis of our current understanding of lunar geology and the original interpretation of some of these features has been questioned [102]. Clearly further work to determine whether or not basaltic lava flows with ages < 1 Gyr exist on the Moon would be desirable.

Probably the main disadvantage of lunar lava flows as preservers of astronomical records within buried regolith layers is the heating of the regolith substrate by the overlying lava when it is emplaced. Detailed studies of this process [103] indicate that the uppermost ~20 cm of regolith covered by a 1 m thick lava flow would likely experience at least partial degassing of implanted volatiles, with thicker lava flows requiring approximately proportional thicker regolith to provide adequate insulation. This sets a lower limit to the thickness of palaeoregoliths able to preserve a good record of solar wind and other implanted volatiles. As regolith production rates are thought to be in the range 1-5 mm Myr$^{-1}$ (where the lower value is the estimated present-day rate and the higher value relates to fresh basaltic surfaces ~3.8 Gyr ago [104,105]), it follows that fresh surfaces would need to be exposed for tens to hundreds of Myr to accumulate regoliths thick enough to shield implanted volatiles from an overlying lava flow. These long regolith accumulation times would lead to a loss of temporal resolution for any astrophysical records they contain. However, it is worth mentioning the recent suggestion [106] that degassing of some mare lavas may result in the formation of metre-thick fragmented layers, 'auto-regoliths', on their surfaces. If present, these would assist in the preservation of volatiles should they subsequently be covered by younger lavas. We also reiterate that a thick palaeoregolith is less of a necessity for the preservation of non-volatile records (e.g. heavy nuclei delivered as SN ejecta and some of the less mobile cosmogenic nuclei produced by GCRs) as these may be preserved within layered basalts lacking interbedded regoliths (such as those shown in Figure 1).[3]

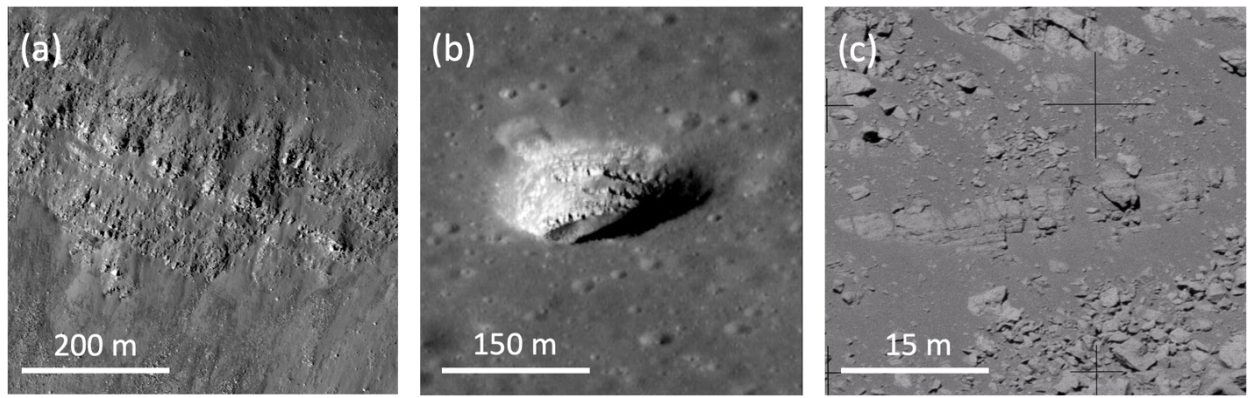

**Figure 1.** (a) Mare basalt stratigraphy in Mare Serenitatis exposed in the wall of the 16 km diameter impact crater Bessel (21.8°N, 17.9°E; LROC image M135073175R/NASA/GSFC/ASU). (b) Similar layering exposed in walls of a collapse pit in Mare Ingenii on the farside (36.0°S, 166.1°E LROC image M184810930L/NASA/GSFC/ASU). (c) Metre-scale basalt layers exposed in the wall of Hadley Rille (26.1°N, 3.6°E) photographed by Apollo 15 astronaut David Scott using a 500mm focal length lens; the outcrop is about 1300m from the camera (NASA image AS15-89-12104).

We illustrate the concept of palaeoregolith preservation and dating with the aid of a prominent lava flow on the surface of Mare Imbrium (Figure 2). Here, a younger lava flow overlies older mare basalts, so we would expect a palaeoregolith layer to be preserved underneath it. This palaeoregolith layer will contain material reaching the surface of the Moon in the time period between the emplacement of the under- and over-lying lavas. Ideally, radiometric dating of returned samples would provide these ages, but as none have yet been obtained from this region we have used standard Crater Size-Frequency Distribution (CSFD) measurements to model the ages of these flows (for details see Electronic Supplementary Information, ESM). Based on our CSFD measurements, the underlying basalts have an absolute model age of $3.30^{+0.04}_{-0.05}$ Gyr, while the over-lying lava flow has an absolute model age of $3.03^{+0.12}_{-0.17}$ Gyr. We therefore hypothesize that sandwiched between these two lava flows there will be a trapped palaeoregolith layer containing records of the solar wind, SCRs, GCRs, and possibly other astronomically valuable records, that reached the surface of the Moon within the ~300 Myr period separating the emplacement of these two lava flows.

---

[3] Although insulation provided by even a thin regolith may be helpful in preserving more volatile GCR products (e.g. $^{36}Cl$, $^{37,39}Ar$) and in preventing the annealing of radiation damage tracks (we thank one of our referees, Gregory Herzog, for this observation).

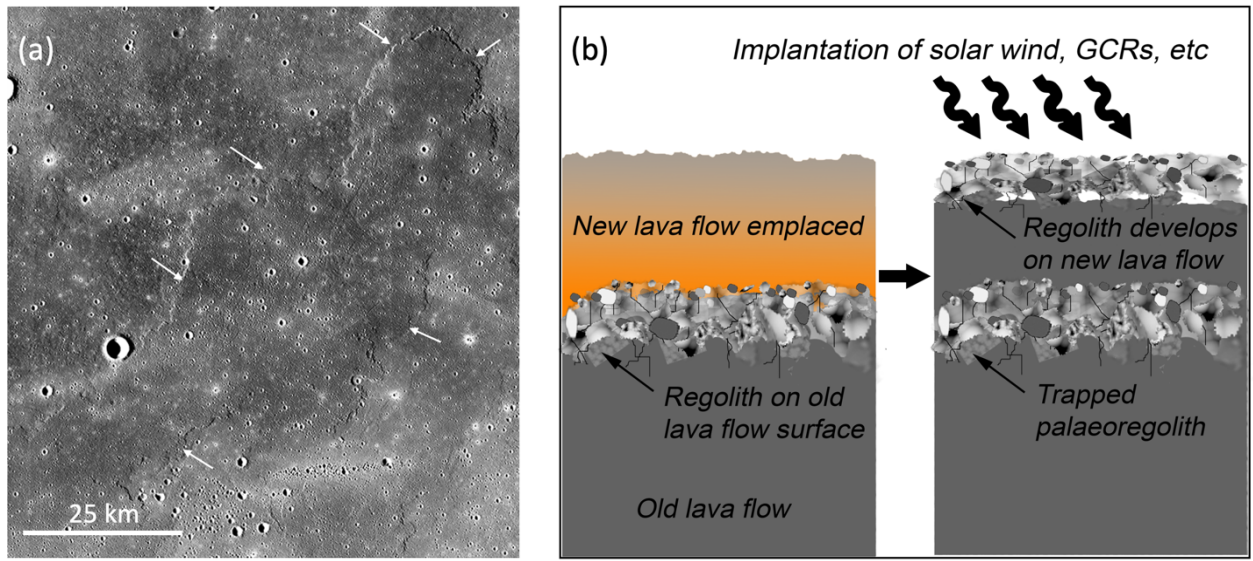

**Figure 2.** (a) A prominent lava flow (arrows) on the surface of Mare Imbrium (31.5°N, 338.0°E; Kaguya Terrain Camera/JAXA). Based on CSFD measurements (see ESM), we obtain an absolute model age of $3.03^{+0.12}_{-0.17}$ Gyr for this lava flow, and an age of $3.30^{+0.04}_{-0.05}$ Gyr for the older lava flows over which it has flowed, so we expect a ~3 Gyr-old palaeoregolith to be trapped between the two (see text). (b) Schematic illustration of the trapping of a palaeoregolith layer by an overlying lava flow; a paleoregolith such as this would be expected to contain solar wind, GCR, and other astrophysical records that were implanted during its time on the surface, meanwhile a new regolith develops on the younger lava flow and captures more recent astrophysical records.

Estimating the extent to which any underlying palaeoregoliths will have been heated by overlying lava flows such as this, as well as drawing up plans for sampling them, requires an estimate of the lava flow thickness. In this case, the CSFD of the underlying surface shows that all measured crater sizes have been resurfaced by the upper lava flow, yielding a minimum flow thickness of ~20m (see ESM). If emplaced as a single lava flow of this thickness the modelling of Rumpf et al [103] would imply that a palaeoregolith below it would need to be several metres thick in order to prevent thermal degassing of solar wind and other trapped volatiles. Such a thick regolith is unlikely to have been generated in the ~300 Myr available given the estimated 1-5 mm/Myr regolith formation rates. On the other hand, there is evidence, for example layers exposed in the wall of Hadley Rille (Figure 1(c)) that some mare basaltic lava flows are built up of multiple thinner (~1 m thick) layers, and if this were the case here it would help reduce the propagation of heat into an underlying regolith [103]; the same would apply if the underlying lava had developed a fragmental 'auto-regolith' on eruption [106]. In any case, as stressed above, non-volatile and some GCR-produced cosmogenic nuclides are much less susceptible to thermal disturbance and are likely to be preserved even under such a relatively thick lava flow.

We stress that we have here merely used these Imbrium lava flows as an example. The comprehensive mapping and dating of mare basalts presented by Hiesinger et al. [100] indicates that there are hundreds of large (≥50 km in size; see figure 17 of [100]) lava flows on the lunar surface spanning the age range ~4.0 – 1.0 Gyr, and doubtless many thousands of smaller examples such as that discussed here, all with the potential to preserve underlying palaeoregolith layers with a correspondingly wide range of ages. Moreover, as noted above, stacks of layered lava flows (Figure 1) have the potential to preserve some astronomically important records (e.g. the GCR flux) in the absence of interleaved palaeoregolith layers. That said, unless evidence for small-scale basaltic volcanism within the last ~1 Gyr is confirmed (e.g., [101]), lava flows are unlikely to preserve astronomical records within this timeframe.

## (b) Pyroclastic deposits

In addition to the effusive eruption of low-viscosity mare basalts, lunar magmatic processes have also resulted in occasional explosive or pyroclastic volcanism [107,108]. The resulting pyroclastic deposits are fine-grained units of basaltic glass fragments mantling surfaces around volcanic vents (Figure 3), with sizes ranging from ~10 km² to ~50,000 km² and thicknesses estimated at several metres [107,109]. McKay [110] has argued that pyroclastic deposits may be the best preservers of palaeoregoliths owing to their relatively gentle mode of emplacement. Moreover, because the small (typically tens of microns in diameter) basaltic glass fragments that make up the deposits would have mostly cooled and solidified before impacting the surface, the buried regolith would not be subject to the thermal disturbances associated with burial by active lava flows.

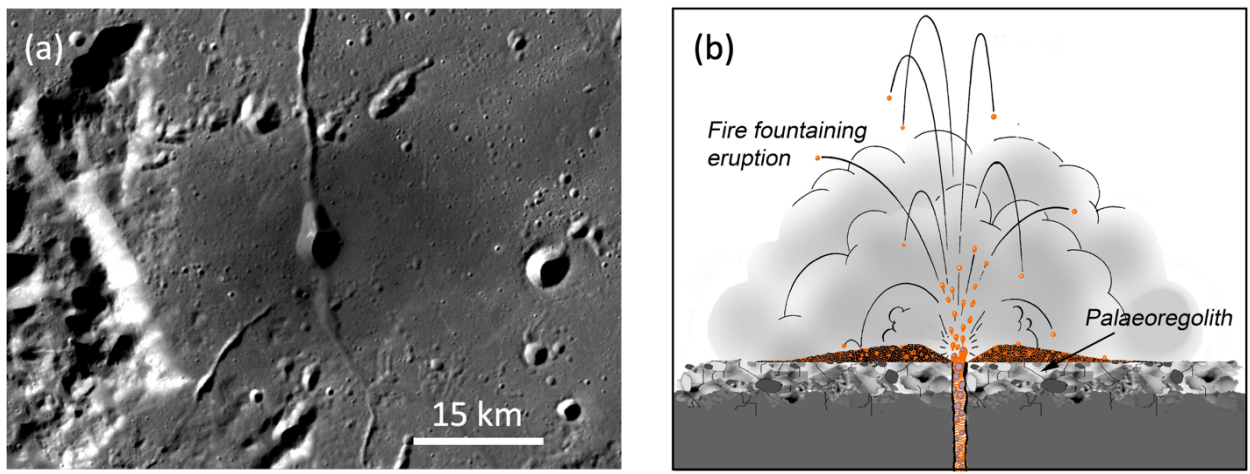

**Figure 3.** (a) Dark pyroclastic materials surrounding a presumed volcanic vent in the Schrödinger Basin on the lunar farside (75.3°S, 139.2°E; LROC image/NASA/GSFC/ASU). (b) Schematic illustration of a pyroclastic eruption covering a pre-existing regolith to preserve a palaeoregolith underneath (adapted with thanks from LPI CLSE Higher Education Resources).

The main disadvantage of pyroclastic deposits in the present context is the ancient, and relatively brief, time periods in which they formed, dating from the main phase of lunar mare volcanism with an estimated age range of 3.8 to 3.2 Gyr [100,110]. While doubtless preserving valuable records of the early Sun, palaeoregoliths buried by currently identified pyroclastic deposits are therefore unlikely to preserve more recent galactic influences on the Solar System. As in the case for basaltic lava flows, a search for more recent pyroclastic deposits would be valuable.

## (c) Ejecta blankets and impact melt deposits

Impact cratering has been continuous throughout lunar history [91], so palaeoregoliths covered by crater ejecta blankets and/or impact melt deposits (Figure 4) have the potential to preserve records from more recent times than those covered by lava flows or pyroclastic deposits. This will be especially important if we seek well-resolved temporal records for the Solar System's most recent orbit around the Galaxy. Dating the emplacement of crater ejecta blankets could be achieved by sampling and radiometrically dating associated impact melt (e.g. [111-113]). Importantly, such ages would be independent of assumptions regarding the GCR and solar wind fluxes that we wish to reconstruct from the underlying palaoregoliths. If the pre-existing regolith was developed on a basaltic lava flow, this too could be sampled and dated, thereby locating the palaeoregolith sandwiched between lava flow and ejecta blanket in a well-constrained time window.

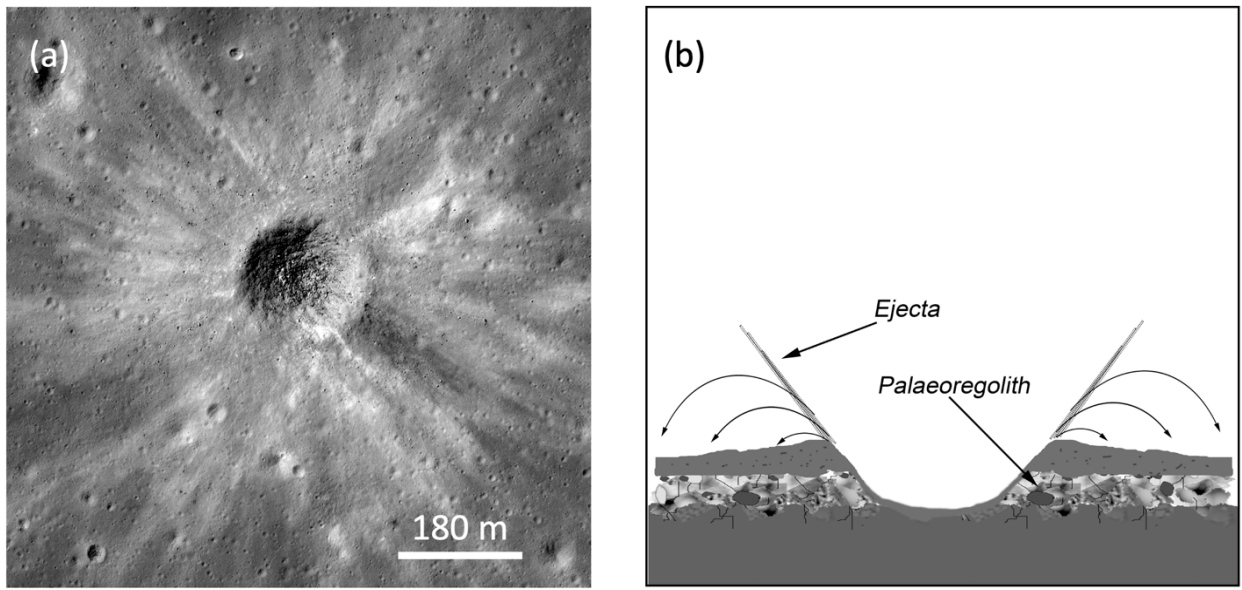

**Figure 4.** (a) An unnamed fresh 185 m diameter impact crater in Mare Nubium (20.9°S, 350.3°E; LROC image M183588912R: NASA/GSFC/ASU); note the prominent ejecta blanket. (b) Schematic illustration of a crater ejecta blanket covering a pre-existing regolith to preserve a an underlying palaeoregolith.

Although impact melt deposits have the potential to preserve palaeoregoliths in the same way as lava flows, they suffer from the same potential disadvantage of thermally disturbing volatile records. The same may be true of palaeoregoliths buried by thick ejecta blankets from large craters which may contain significant volumes of impact-heated materials [114]. On the other hand, the ejecta of small craters is not thought to be at a high temperature when emplaced, so the main uncertainty in their value as preservers of buried palaeoregoliths is the degree to which the pre-existing surface is mechanically disturbed in the process. As the ejecta is mostly emplaced ballistically [115], this is likely to depend on the size of the impact and the distance from it, and may be quite variable. It is also likely to disturb some records more than others, with very surficial records (e.g. solar wind, pick-up ions and SN ejecta) being more disturbed than deeper-lying cosmogenic nuclides produced by GCR interactions. Further work on the mechanical and thermal effects of impact ejecta emplacement would be desirable.

# 5. Locating and accessing the record

Gaining access to these astronomical records will present considerable technical challenges. There are two main aspects: identifying the most promising locations where such records are likely to be preserved, and accessing and sampling these locations.

**(a) Locating the records**

We have argued that the astronomical records we seek will be preserved in sub-surface layers, such as buried palaeoregoliths or lava flows, that were once exposed at the lunar surface. Practical considerations suggest that an initial search must be based on surface features accessible to remote-sensing techniques that are indicative of the likely presence of suitable sub-surface deposits. Examples include the geological mapping and dating of surface lava flows (e.g. [100]), pyroclastic deposits (e.g. [107]), and the ejecta blankets of small Copernican-aged craters. In addition, high resolution images of the lunar surface have revealed multiple locations where sub-surface layers out-crop in the walls of rilles, craters, and collapse pits ([99]; Figure 1). Studies of areas where small impact craters have penetrated overlying materials to reveal sub-surface boundaries may also help identify suitable locations [116], as would orbital ground penetrating radar measurements [117]. Ultimately, it will probably be necessary to visit a sub-set of identified localities with robotic or human explorers employing geophysical techniques, such as ground penetrating radar [118,119] or refraction seismology [120], to confirm the existence of suitable sub-surface deposits and to assess the practicalities of sampling them.

**(b) Accessing the records**

An optimal architecture for accessing and sampling sub-surface deposits would provide for the following capabilities [50,51]:

- Ability to deploy equipment at a wide (preferably global) range of locations on the lunar surface
- Surface mobility, ideally with a range of several tens of km around a given landing site (for example, this would permit access to the boundaries of lava flows having a wide range of ages; see, e.g., the mare basalt maps provided by [100])
- Detection of sub-surface palaeoregolith deposits (e.g. using ground penetrating radar or active seismic profiling [119,120])
- Access and sampling of outcrops on steep slopes such as crater walls or entrances to collapse pits (e.g. [121])
- Drilling from 10s of metres to perhaps ~100 metre depths (for a review of suitable planetary drilling technology see [122,123])
- Return of samples to Earth for analysis; the quantity required will depend on the number and types of sites (e.g. palaeoregolith layers) sampled, but based on the analysis of Shearer et al. [124] we estimate this to be ~100 kg for each exploration mission or sortie (which would presumably visit multiple individual localities; compare with the 110 kg returned by the Apollo 17 mission).

Although some of these capabilities might be achievable with suitable designed robotic missions (e.g. [121]), as argued elsewhere (e.g. [1,2]) large scale exploratory activities such as these would be enhanced by renewed human operations on Moon (Figure 5). Especially enabling would be the creation of one or more permanently, or semi-permanently, occupied scientific research stations on the lunar surface, as exemplified by the 'Moon village' concept advocated by the Director General of the European Space Agency [125]. The creation of such an outpost would offer significant opportunities by providing a scientific infrastructure on the lunar surface, just as human outposts in Antarctica facilitate research activities across multiple scientific disciplines [126-128].

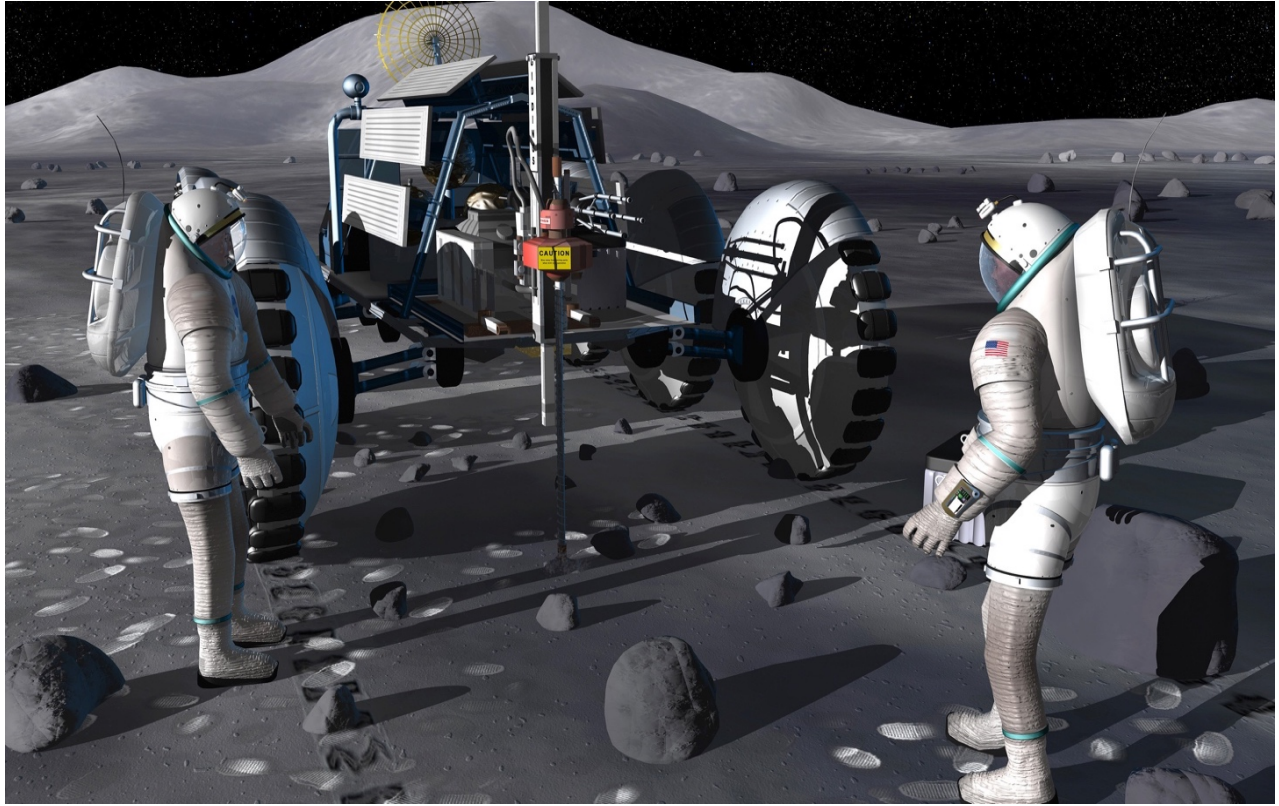

**Figure 5.** Artist's concept of astronauts supervising a lunar drilling system. Such a capability would permit access to the sub-surface, for example to extract palaeoregolith samples containing ancient solar wind and GCR records, and is an example of how astrophysics, among other sciences, could benefit from establishing a human-tended research infrastructure on the Moon (image credit: NASA).

## 6. Conclusion

The Moon is likely to preserve a rich historical record of astrophysical processes relevant to understanding the evolution of the Sun and its changing galactic environment. In order to access these records, it will be necessary to collect samples from sub-surface strata that were directly exposed to the space environment at known times in the past and for known durations. Such geological records undoubtedly exist on the Moon but accessing them will require a greatly expanded programme of lunar exploration. Ideally, this will include the eventual establishment of Antarctic-style research stations to support large-scale exploration activities. Such a research infrastructure would also support a wide range of other scientific activities on the Moon [2,4], including, in the present context, observational astronomy and astrophysics.

## Additional Information

**Data accessibility.** Additional data can be found in the associated electronic supplementary material.

**Authors' contributions.** IAC conceived the paper and drafted most of the manuscript. KHJ provided most of the expertise relating to solar wind in the lunar regolith and produced the diagrams illustrating palaeoregolith preservation. JHP and HH performed the CSFD measurements of the Imbrium lava flows. All authors provided intellectual content and critically reviewed the whole manuscript prior to submission.

**Competing interests.** We declare that we have no competing interests.

**Funding.** We acknowledge STFC grant ST/M001253/1 and Royal Society grant UF140190 to KHJ and Leverhulme Trust grant RPG-2019-222 to KHJ and IAC.

**Acknowledgements.** We thank our two referees, Gregory Herzog and the other anonymous, for helpful comments that have improved the manuscript. We thank Joe Silk for drawing our attention to possible geological recorders of astrophysical sources of neutrinos and dark matter particles. IAC also thanks Ellie Ganpot for helpful discussions.

# Electronic Supplementary Information

Crater Size-Frequency Distribution (CSFD) Measurements

1. Method

1.1 Absolute Model Ages

Crater Size-Frequency Distribution (CSFD) measurements are a widely used method to derive relative and absolute model ages of planetary surfaces using remote sensing data [e.g., 1-8]. We used CraterTools in ArcGIS developed by Kneissl et al. [9] to measure the diameters and frequencies of primary craters and the sizes of the respective counting area. For the analyses of the measured CSFDs, CraterStats II, developed by Michael and Neukum [10], has been utilized. To derive relative and absolute ages from the CSFD measurements, the lunar production (PF) and chronology (CF) functions have to be known. For our study, we used the PF of Neukum et al. [5] which is given by an eleventh-order polynomial:

$$\log(N_{cum}) = a_0 + \sum_{j=1}^{11} a_j (\log(D))^j \tag{1}$$

where $N_{cum}$ is the cumulative number of craters per km², $a_0$ represents the time over which a certain unit has been exposed to the meteorite bombardment, and $D$ is the crater diameter in km [2,5,11].

We also applied the empirically derived lunar chronology function (CF) of Neukum et al. (2001), which is given by the following equation:

$$N_{cum}(D \geq 1 \text{ km}) = 5.44 \times 10^{-14}[\exp(6.9 \times t) - 1] + 8.38 \times 10^{-4} t \tag{2}$$

In this equation, $t$ is the crater accumulation time (crater retention age) in Ga, and $N_{cum}$ is again the cumulative number of craters per km².

In order to achieve reliable relative and absolute model ages, we used spectrally, topographically, and morphologically homogenous counting areas and excluded areas with secondary crater clusters [e.g. 12-17].

To determine relative ages of the mapped geologic units, we calculated the sizes of the count areas, counted all craters apart from secondary and endogenic craters, measured their diameters measured, and fitted the data with the PF of Neukum et al. [5]. The derived crater frequency of craters larger or equal 1 km were linked to the CF of Neukum et al. [5] to obtain absolute model ages (AMAs). We used Poisson statistics, introduced by Michael at al. [18], to fit the CSFDs and to derive AMAs. This approach uses a probability density function, which directly links the chronology function with the measured crater density of the surface. To evaluate our CSFD measurements we used the randomness analyses described by Michael et al. [19]. This method determines the level of clustering by measuring the mean 2nd-closest neighbor distances (M2CND) of every used crater size bin.

1.2 Flow Thickness

The flow thickness can be estimated based on the rim height of flooded or partly flooded impact craters (e.g., Hiesinger et al. [20]). The relation between the rim height and the crater diameter for simple craters < 15 km is given by Pike [13]:

$$R_e = 0.036 \, D_r^{1.014} \tag{3}$$

where $R_e$ is the rim height (km) and $D_r$ is the crater diameter (km).

2 Results

We derived absolute model ages for two lava flows within the Imbrium basin (see Figure 2 of the main paper). The two flows are separated by a distinct morphologic boundary. Based on stratigraphic observations on Kaguya images the topographically higher flow seems to have overflown the lower flow. However, as only the upper flow shows a clear flow front, and spectral differences are low, the overflown surface cannot be separated into individual lava flows. Thus, the relative stratigraphic relationship between the upper and the lower flow could be more complex than visible with the available data sets. To determine the relative and absolute age difference of the two flows we performed CSFD measurements of the two flows, respectively. The counting area locations of these measurements are shown in Supplementary Figure 1.

Based on our CSFD measurements, the lower flow shows an AMA of 3.3 +0.04/-0.05 Ga (Supplementary Figure 2). The upper flow shows an AMA of 3.03 +0.12/-0.17 Ga (Supplementary Figure 3) being ~300 Ma younger than the lower flow. Since the error bars of both AMAs are relatively small, this age difference seems to be reliable. The CSFD of the lower flow shows that all measured crater sizes have been resurfaced by the upper flow. The largest measured crater shows a diameter of 473 m and allows us to estimate the minimum flow thickness of the upper flow. Based on the rim height / crater diameter relation for lunar simple craters determined by Pike [13] this translates into a minimum flow thickness of 18.6 m.

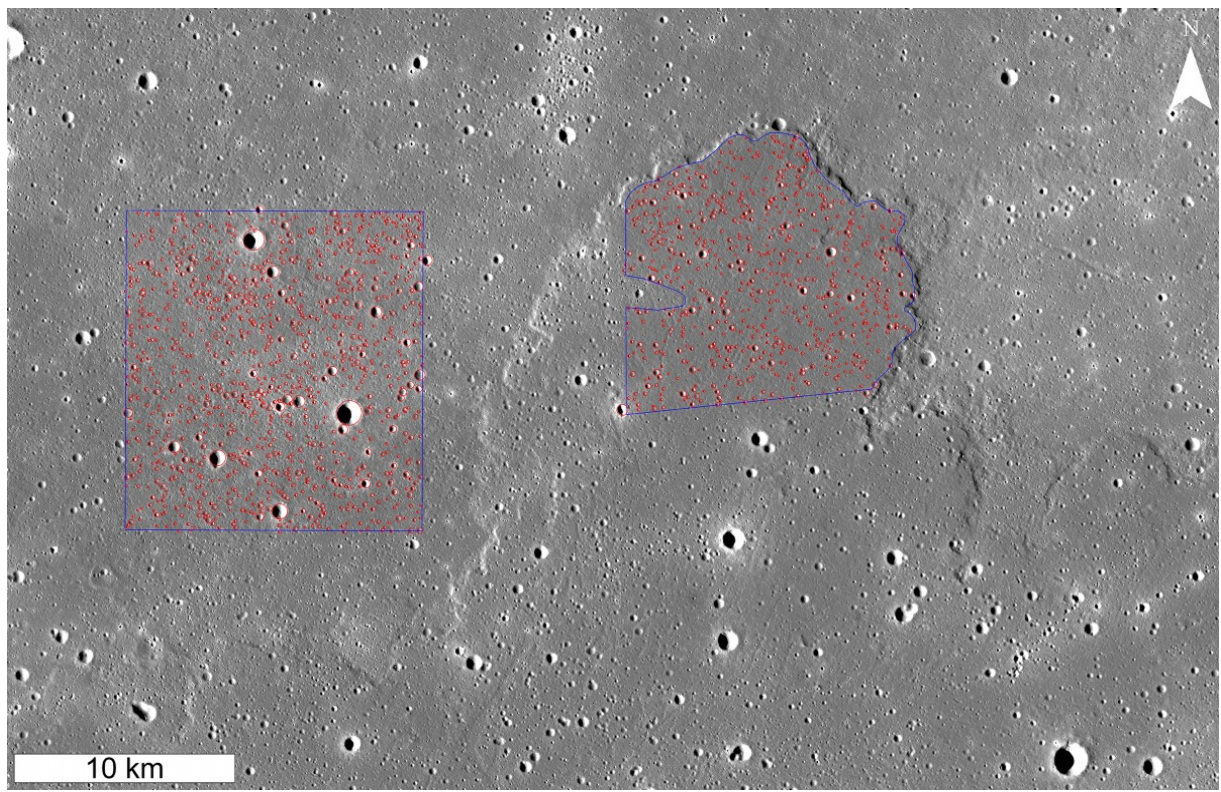

**Supplementary Figure 1:** Counting areas of the low (left) and the upper (right) flow in blue. The measured craters are marked in red. Base map is the 10 m/pixel Kaguya Terrain Camera (TC) mosaic (Image centre at 32.3°N and 22.1° W).

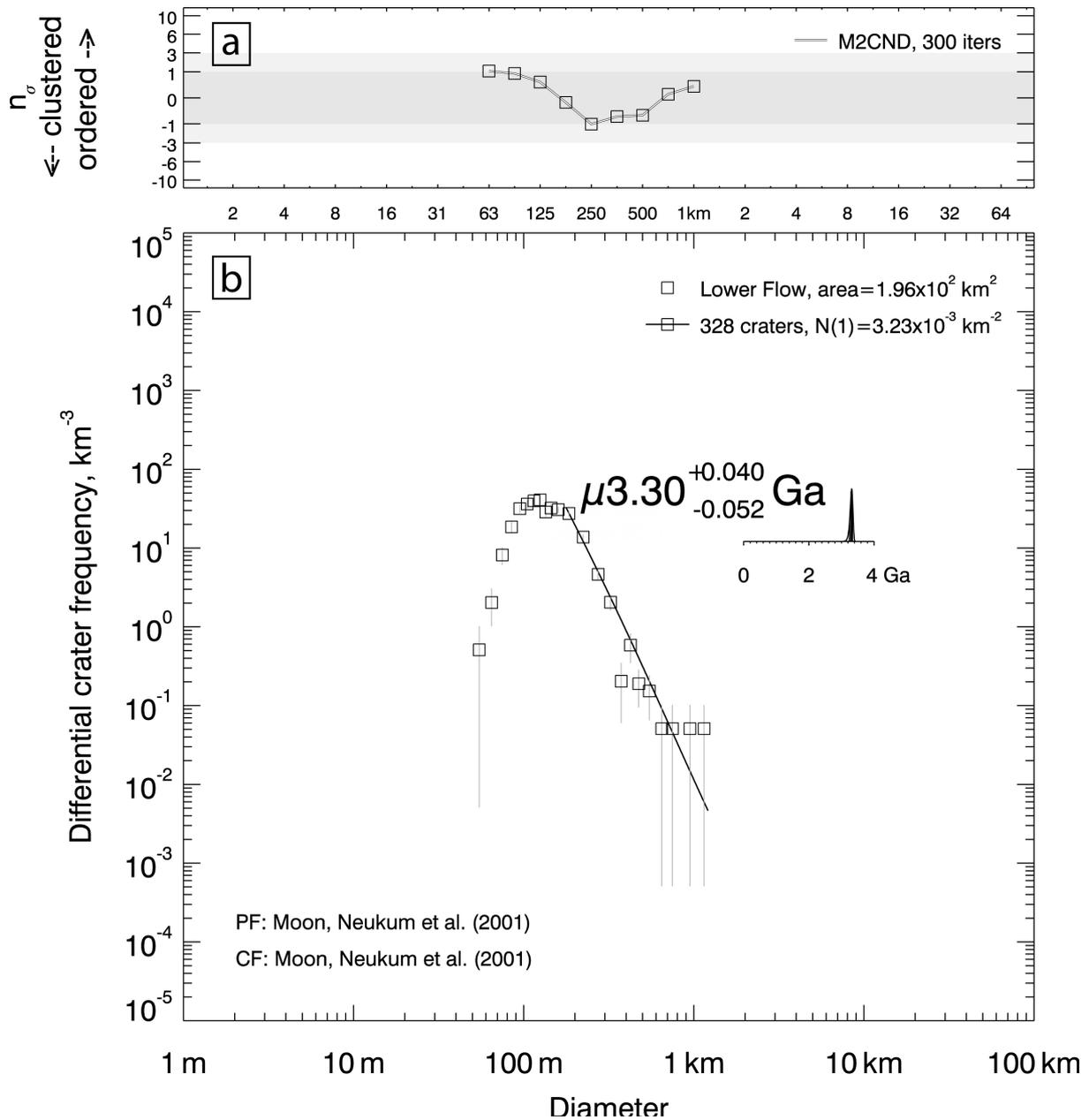

**Supplementary Figure 2.** CSFD analysis of the lower lava flow. **(a)** Summary plot of the M2CND randomness analyses plotted in standard deviations (σ), above or below the Monte Carlo-derived mean; as most of the crater size bins are between +1 and -1 standard deviations, the influence of clustering due to secondary cratering is negligible. **(b)** CSFDs of the lower flow, shown in a differential plot; the calculated AMA (μ) and error bars come along with a plot of the uncertainty as a probability density function, marked at the 50 and 50 ±34 percentiles.

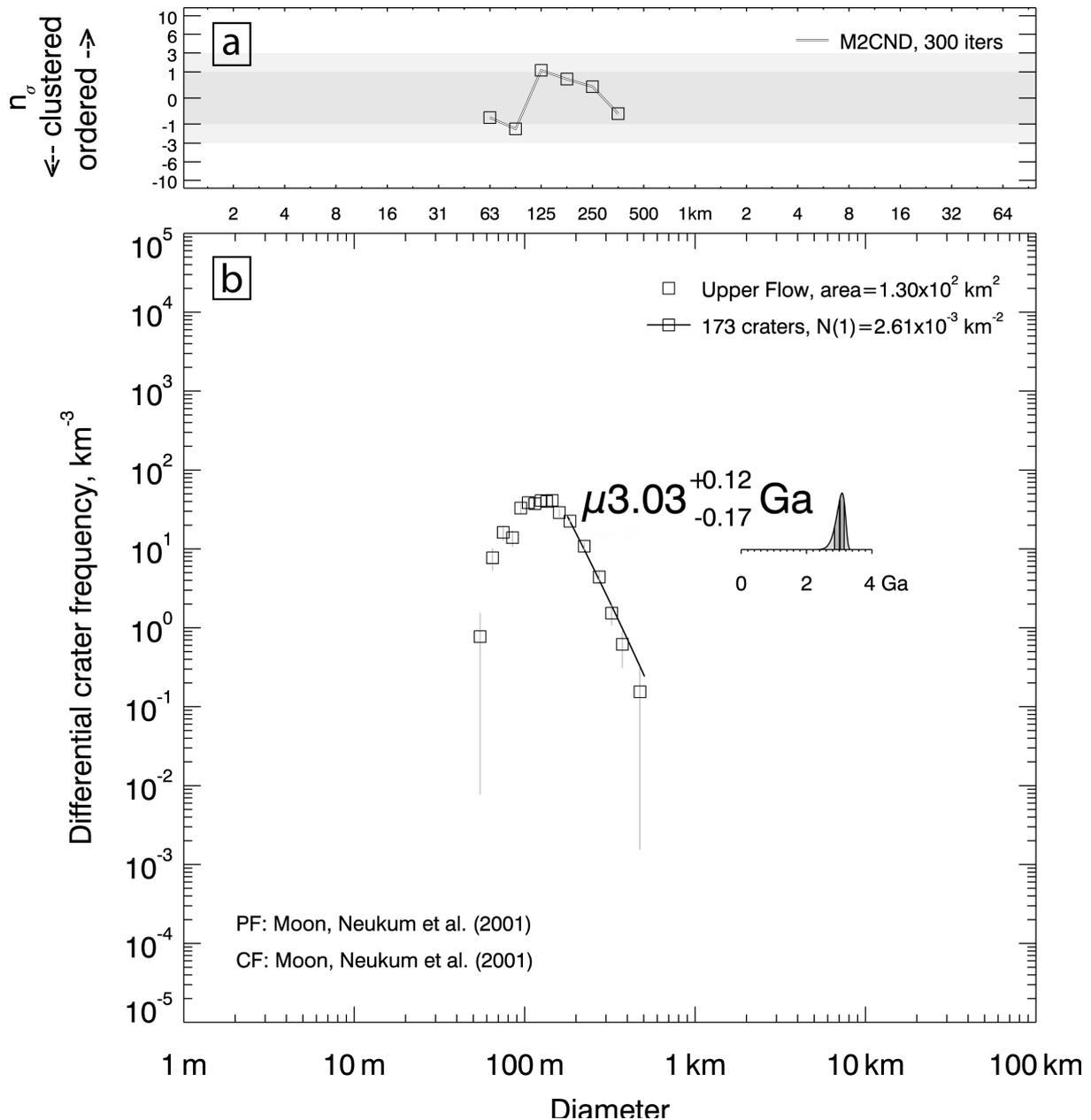

**Supplementary Figure 3.** CSFD analysis of the upper lava flow. **(a)** Summary plot of the M2CND randomness analyses plotted in standard deviations (σ), above or below the Monte Carlo-derived mean; as most of the crater size bins are between +1 and -1 standard deviations, the influence of clustering due to secondary cratering is negligible. **(b)** CSFDs of the upper flow, shown in a differential plot; the calculated AMA (μ) and error bars come along with a plot of the uncertainty as a probability density function, marked at the 50 and 50 ±34 percentiles.

## References

1. Crater Analysis Techniques Working Group (Arvidson, R.E., Boyce, J., Chapman, C.R., Cintala, M.J., Fulchignoni, M., Moore, H., Neukum, G., Schultz, P.,